\documentclass[pageno]{jpaper}

\usepackage{mathptmx} 
\usepackage{fancyhdr}
\usepackage[normalem]{ulem}
\usepackage[keeplastbox]{flushend}
\usepackage[english]{babel}
\usepackage[utf8]{inputenc}
\usepackage{algorithm}
\usepackage{float}
\usepackage[noend]{algpseudocode}
\usepackage{textcomp}
\usepackage[dvipsnames]{xcolor}
\usepackage{xspace}
\usepackage{subcaption}
\usepackage{booktabs}
\usepackage{enumitem}
\usepackage[noadjust]{cite}
\usepackage{graphicx}
\usepackage{pifont}
\usepackage{setspace}
\usepackage{mwe}
\usepackage{multirow}
\usepackage[compact]{titlesec}
\usepackage{amsmath}
\usepackage{tikz}

\newcommand{\mech}{{SIMDRAM}\xspace} 

\newcommand{\circled}[1]{\tikz[baseline=(char.base)]{\node[shape=circle,draw,inner sep=0pt,fill=black, text=white] (char) {#1};}}

\makeatletter

\renewcommand\newblock{\hskip .11em\@plus.33em\@minus.07em}
\makeatother

\begin{document}

\bstctlcite{IEEEexample:BSTcontrol}

\title{Methodologies, Workloads, and Tools for Processing-in-Memory: \\
Enabling the Adoption of Data-Centric Architectures}
\newcommand{\affilETH}{$^\diamond$}
\newcommand{\affilCMU}{$^\ddag$}
\newcommand{\affilUIUC}{$^\odot$}
\newcommand{\affilSFU}{$^\star$}

\author{
        \vspace{-15pt}\\ \scalebox{1.0}{Geraldo F. Oliveira}\affilETH~\qquad%
        \scalebox{1.0}{Juan Gómez-Luna}\affilETH~\qquad%
        \scalebox{1.0}{Saugata Ghose}\affilUIUC~\qquad%
        \scalebox{1.0}{Onur Mutlu}\affilETH%
    \\%
    \vspace{-10pt}\\%
        \it\normalsize \affilETH ETH Z{\"u}rich  \qquad  \affilUIUC University of Illinois Urbana-Champaign
    \vspace{-20pt}%
}

\date{}
\maketitle

\section{Motivation \& Problem}
\vspace{-5pt}

The increasing prevalence and growing size of data in modern applications
{have} led to high costs for computation in traditional {processor-centric computing} systems.
Moving large volumes of data between memory devices (e.g., DRAM) and {computing elements (e.g., CPUs, GPUs)} across 
bandwidth-limited memory channels can consume more than
60\% of the total energy in modern systems~\cite{mutlu2019, boroumand2018google}.
To mitigate these costs, {the \emph{processing-in-memory} (PIM)~\cite{ghoseibm2019, mutlu2020modern,oliveira2021pimbench,pim-book,mutlu2019,mutlu2019enabling,mutlu2015research,mutlu2013memory} paradigm moves} computation closer to where the 
data resides, reducing (and in some cases eliminating) the
need to move data between memory and the processor.

There are two main approaches to PIM~\cite{mutlu2020modern}:
(1)~processing-near-memory (\emph{PnM})~\cite{ahn2015scalable, nai2017graphpim, boroumand2018google, lazypim, top-pim, gao2016hrl, kim2018grim, drumond2017mondrian, santos2017operand, NIM, PEI, gao2017tetris, Kim2016, gu2016leveraging, boroumand2019conda, hsieh2016transparent, cali2020genasm, NDC_Micro_2014, farmahini2015nda,loh2013processing,pattnaik2016scheduling,akin2016data, hsieh2016accelerating,babarinsa2015jafar,lee2015bssync, devaux2019true,boroumand2021mitigating,boroumand2021google,boroumand2022polynesia,boroumand2021polynesia, amiraliphd,ghiasi2022genstore,gomez2021benchmarkingcut,gomez2021benchmarking,gomez2022benchmarking,besta2021sisa,syncron,fernandez2020natsa,singh2020nero,rezaei2020nom,singh2019napel,lee2022isscc,kwon202125,lee2021hardware,ke2021near,niu2022184qps,giannoula2022sparsep,shin2018mcdram,cho2020mcdram,Sparse_MM_LiM,azarkhish2016logic,azarkhish2017neurostream,guo20143d,de2018design,akin2014hamlet,huang2020heterogeneous,dai2018graphh,liu2018processing,tsai:micro:2018:ams,denzler2021casper,gu2020ipim,asghari2016chameleon,IRAM_Micro_1997,Near-Data,C_RAM_1999,CASES_MVX,DRAMA_CAL_2014,Asghari-Moghaddam_2016,Xi_2015}, where PIM logic is added to the same die as memory or to the logic layer of 3D-stacked memory~{\cite{HMC2,HBM,lee2016simultaneous}}; and (2)~processing-using-memory (\emph{PuM})~\cite{Chi2016, Shafiee2016, seshadri2017ambit, seshadri2019dram, li2017drisa, seshadri2013rowclone, seshadri2016processing, deng2018dracc, xin2020elp2im, song2018graphr, song2017pipelayer,gao2019computedram, eckert2018neural, aga2017compute,dualitycache,besta2021sisa,seshadri.arxiv16,seshadri.bookchapter17,seshadri2018rowclone,seshadri2015fast,pinatubo2016,ferreira2021pluto,imani2019floatpim,he2020sparse}, which uses the operational principles of memory cells to perform computation (for example, by exploiting DRAM's analog operation to perform bulk bitwise AND, OR, and NOT logic operations~{\cite{seshadri2017ambit, seshadri2015fast, seshadri2019dram,seshadri2016processing,seshadri.bookchapter17,seshadri.arxiv16}}).

{Many works from {academia}~{\cite{PEI, ahn2015scalable, nai2017graphpim, song2018graphr, zhang2018graphp, angizi2019graphs, angizi2019graphide, zhuo2019graphq,gao2017tetris, Kim2016,Shafiee2016,Chi2016,boroumand2021mitigating,amiraliphd,kim2018grim,NIM, cali2020genasm,drumond2017mondrian, santos2017operand, lazypim, boroumand2019conda,hsieh2016accelerating,boroumand2021polynesia,boroumand2022polynesia,gu2016leveraging,boroumand2021google,kim2019d,kim2018dram,seshadri2017ambit,li2017drisa,seshadri2013rowclone, wang2020figaro, chang2016low, rezaei2020nom, seshadri2015fast,boroumand2018google,cali2022segram}} and {industry}~\cite{lee2022isscc,kwon202125,lee2021hardware,ke2021near,niu2022184qps,devaux2019true,gomez2021benchmarking,gomez2021benchmarkingcut,gomez2022benchmarking} have shown the benefits of PnM and PuM for a wide range of workloads from different domains. However, fully adopting PIM in commercial systems is still very challenging due to {the} lack of {tools and} system support for PIM architectures across the computer architecture stack~\cite{mutlu2020modern}, which includes:
\emph{(i)} {workload characterization} methodologies and benchmark {suites targeting} PIM architectures;
\emph{(ii)} frameworks that can facilitate {the implementation of complex operations and algorithms using the underlying} PIM primitives {(e.g., simple PIM arithmetic operations~\cite{PEI}, bulk bitwise Boolean in-DRAM operations~\cite{seshadri2017ambit, seshadri2019dram, gao2019computedram})};
\emph{(iii)} compiler support and compiler optimizations targeting PIM architectures;
\emph{(iv)} operating system support for PIM-aware virtual memory, memory management, data allocation {and} mapping; and \emph{(v)} efficient data coherence and consistency mechanisms.}

{Our \emph{goal} in this work is to {provide} {tools and} system support for PnM and PuM architectures, aiming to ease the adoption of PIM in current and future systems. With this goal in mind, we address two limitations of prior works related to {\emph{(i)}} identifying and characterizing workloads suitable for PnM offloading and {\emph{(ii)}} enabling complex operations in PuM architectures. First, we develop a methodology, called \emph{DAMOV}, that identifies sources of data movement bottlenecks in applications and associates such bottlenecks with PIM suitability. Second, we propose an end-to-end framework, called \emph{SIMDRAM}, that enables the implementation of complex in-DRAM operations transparently {to} the {programmer}.}

\section{DAMOV: Identifying and Characterizing Data Movement Bottlenecks}
\vspace{-5pt}

DAMOV introduces the first rigorous methodology to characterize memory-related data movement bottlenecks in modern workloads and the first {benchmark suite for data movement related studies.} {We} develop a new methodology to correlate application characteristics with the \emph{primary} sources of data movement bottlenecks and to determine the potential benefits of three example data movement mitigation mechanisms: (1) a deep cache hierarchy, (2) a hardware prefetcher, and (3) a general-purpose PnM architecture. 

\begin{figure}[ht]
    \centering
    \includegraphics[width=\linewidth]{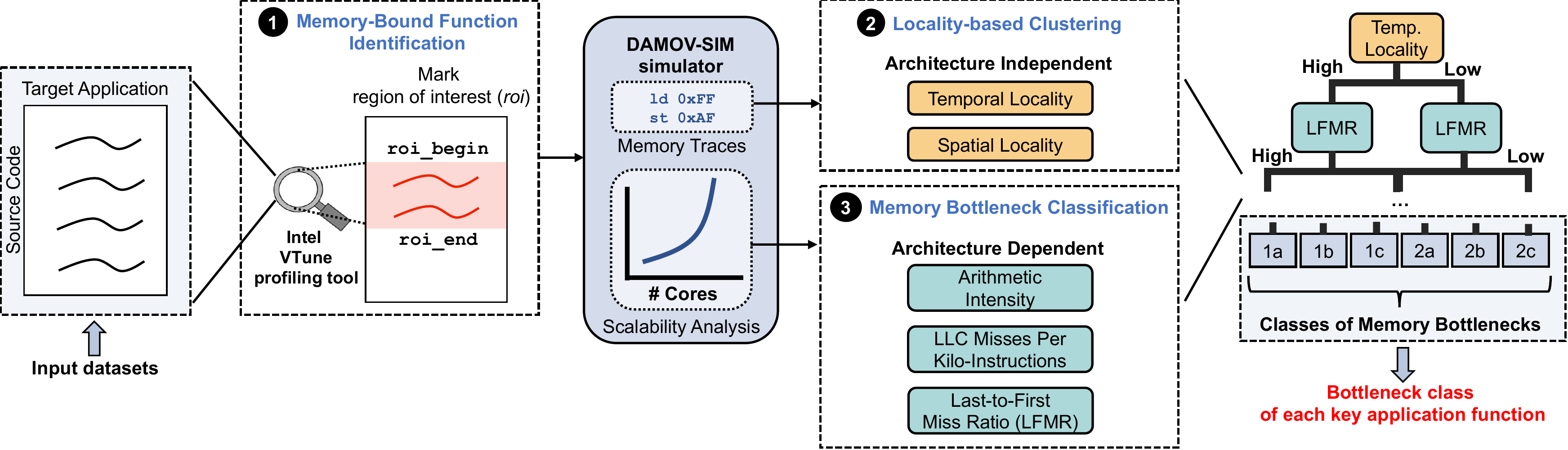}
    \caption{Overview of our three-step workload characterization methodology.}
    \label{figure_methodology}
    \vspace{-10pt}
\end{figure}

Our methodology has three steps. In \textit{Step~1} {(\circled{1} in Figure~\ref{figure_methodology})}, we use a hardware profiling tool~{\cite{vtune}} to identify memory-bound functions across applications. In \textit{Step~2} {(\circled{2})}, we use an architecture-independent profiling tool~{\cite{weinberg2005quantifying,shao2013isa}} to collect metrics that provide insights about the memory access behavior of each function. In \textit{Step~3} {(\circled{3})}, we collect architecture-dependent metrics and analyze the performance and energy of each function on our three data movement mitigation mechanisms. By combining the three steps, we systematically classify the leading causes of data movement bottlenecks in an application or function into different bottleneck classes. 

\vspace{-10pt}

Using this new methodology, we characterize 345~applications from a wide range of domains. Within these applications, we find (and fully characterize) 144~functions that are memory-bound and significantly contribute to the overall execution time. These functions are the core of our data movement benchmark suite, called DAMOV~\cite{damov}. Our analyses reveal six new insights about the sources of data movement bottlenecks and their relation to PnM:
\begin{enumerate}[noitemsep, leftmargin=*, topsep=0pt]
    \item Applications with high last-level cache misses per kilo-instruction (MPKI) and low temporal locality are \emph{DRAM bandwidth-bound}. These applications benefit from the large memory bandwidth available to the PnM system. 
    
    \item Applications with low last-level cache MPKI and low temporal locality are \emph{DRAM latency-bound}. These applications do \emph{not} benefit from L2/L3 caches. The PnM system improves performance and energy efficiency by sending L1 misses directly to DRAM. 
    
     \item A second group of applications with low LLC MPKI and low temporal locality are \emph{bottlenecked by L1/L2 cache capacity}. These applications benefit from the PnM system at low core counts. However, at high core counts (and thus larger L1/L2 cache space), the caches capture most of the data locality in these applications, decreasing the benefits the PnM system provides. We make this observation using a \emph{new} metric that we develop, called \emph{last-to-first miss-ratio (LFMR)}, which we define as the ratio between the number of LLC misses and the total number of L1 cache misses. We find that this metric accurately identifies how efficient the cache hierarchy is in reducing data movement.
     
    \item Applications with high temporal locality and low LLC MPKI are \emph{bottlenecked by L3 cache contention} at high core counts. In such cases, the PnM system provides a cost-effective way to alleviate cache contention over increasing the L3 cache capacity.
    
    \item Applications with high temporal locality, low LLC MPKI, and low arithmetic instruction (AI) are bottlenecked by the \emph{L1 cache capacity}. The three candidate data movement mitigation mechanisms achieve similar performance and energy consumption for these applications.
    
    \item  Applications with high temporal locality, low LLC MPKI, and high AI are \emph{compute-bound}. These applications benefit from a deep cache hierarchy and hardware prefetchers, but the PnM system degrades their performance.
\end{enumerate}

We publicly release our 144~representative data movement bottlenecked functions from 74 applications as the first open-source benchmark suite for data movement, called DAMOV benchmark suite, along with the complete source code for our new characterization methodology {and simulator}~\cite{damov}. {For more information on our extensive data movement bottleneck characterization and on our DAMOV benchmark suite, along with our detailed contributions {(including four use cases of our benchmark suite)}, please refer to our full paper~\cite{oliveira2021pimbench,deoliveira2021arxiv}.}  

\section{SIMDRAM: Enabling Complex {Operations using DRAM}}

\vspace{-8pt}
{A common approach for PuM architectures is to make use of bulk bitwise computation. Many widely-used data-intensive applications (e.g., databases, neural networks, graph analytics) heavily rely on a broad set of simple (e.g., AND, OR, XOR) and complex (e.g., equality check, multiplication, addition) bitwise operations. Ambit~\cite{seshadri2017ambit, seshadri2015fast, seshadri2019dram,seshadri2016processing,seshadri.bookchapter17,seshadri.arxiv16}, an in-DRAM PuM accelerator, proposes exploiting DRAM's analog operation to perform bulk bitwise {majority-of-three (MAJ) computation, which can be manipulated to perform} AND, OR, and NOT logic operations. Inspired by Ambit, many prior works have explored DRAM {and emerging non-volatile memory (NVM)~\cite{lee2009architecting,
qureshi2009scalable,
lee2010phase,
lee2010phasecacm,
kultursay2013evaluating,
zhou2009durable,
wong2010phase, 
meza2012case, 
meza2013case, 
song2020improving,
song2021aging, 
song2019enabling,
bock2011analyzing,
burr2008overview,
du2013bit,
jiang2012fpb,
jiang2013hardware,
kannan2016energy,
qureshi2011pay,
qureshi2010improving,
qureshi2010morphable,
sebastian2017temporal,
wang2015exploit,
yue2013accelerating,
zhou2012writeback,
zhou2013writeback,
yoon2013techniques,
DAC-2009-DhimanAR,
wang2013low,
chen2010advances,
diao2007spin,
hosomi2005novel,
raychowdhury2009design,wong2012metal,yang2013memristive,bondurant1990ferroelectronic,yoon2014efficient}} designs that are capable of performing in-memory bitwise operations~\cite{angizi2019graphide, angizi2018imce, ali2019memory, pinatubo2016, gao2019computedram, xin2020elp2im,li2018scope}.
However, a major shortcoming prevents these proposals from becoming widely applicable: 
they support only basic operations (e.g., Boolean operations, addition) and fall short on flexibly supporting new and more 
complex operations. Our \emph{goal} is to design a framework that aids the adoption of processing-using-DRAM by efficiently implementing complex operations and providing the flexibility }to support new desired operations.

\begin{figure}[ht]
\begin{subfigure}{\linewidth}
    \centering
    \includegraphics[width=\textwidth]{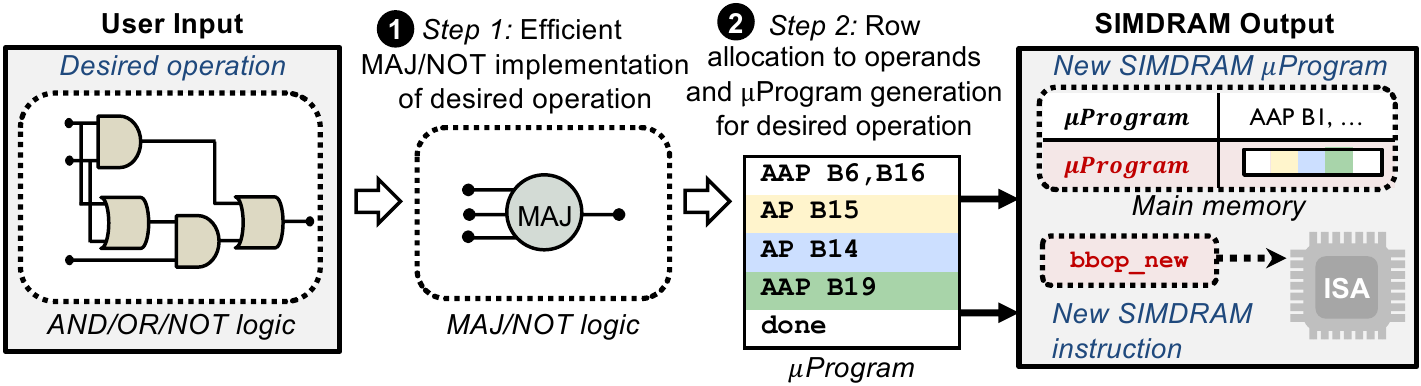}%
    \vspace{-4pt}
    \caption{SIMDRAM Framework: Steps 1 and 2}
    \label{fig_framework_1_2}
\end{subfigure}
\par\bigskip 
\begin{subfigure}{\linewidth}
  \centering
    \includegraphics[width=\textwidth]{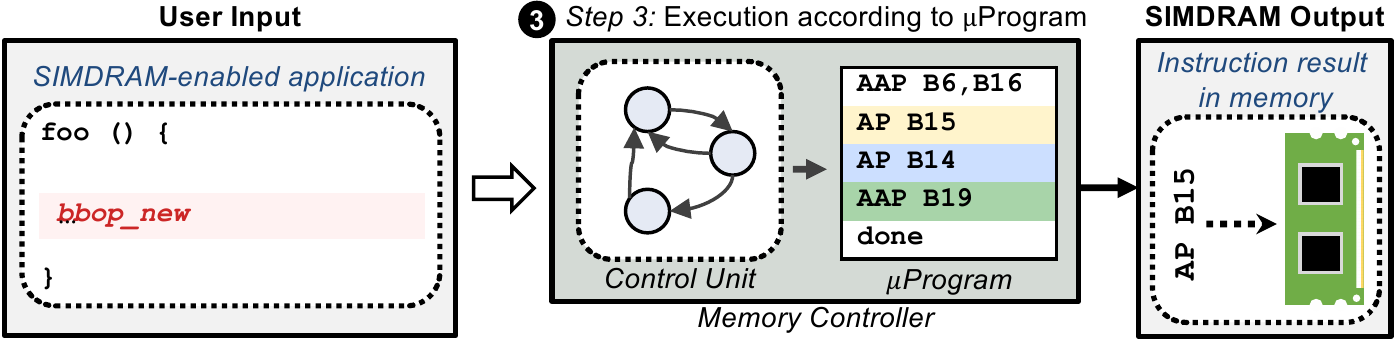}%
    \vspace{-4pt}
    \caption{SIMDRAM Framework: Step 3}
    \label{fig_framework_3}
\end{subfigure}
\caption{Overview of the \mech framework.}
\label{fig:figoverview}
\end{figure}

{To this end, we propose SIMDRAM,} the first end-to-end framework for processing-using-DRAM. At its core, we build the SIMDRAM framework around a DRAM substrate that enables two previously-proposed techniques: (1) vertical data layout in DRAM to support bit-shift operations, and (2) majority-based logic. SIMDRAM consists of three key steps{, illustrated in Figure~\ref{fig:figoverview},} to enable a desired operation in DRAM: (1)~building an efficient MAJ/NOT-based representation of the desired operation, (2)~mapping the operation input and output operands to DRAM rows and to the required DRAM commands that produce the desired operation, and 
(3)~executing the operation. We briefly describe these steps.

\vspace{-8pt}
The goal of the first step {(\circled{1} in Figure~\ref{fig_framework_1_2})} is to use logic optimization to minimize the number of DRAM row activations, and therefore the compute latency required to perform a specific operation. Accordingly, for a desired computation, the first step is to derive its \emph{optimized} MAJ/NOT-based implementation from its AND/OR/NOT-based implementation.

The second step {(\circled{2} in Figure~\ref{fig_framework_1_2})} translates the MAJ/NOT-based implementation into DRAM row activations. This step includes (1)~mapping the operands to the designated rows in DRAM, and (2)~defining the sequence of DRAM row activations that are required to perform the computation. \mech chooses the operand-to-row mapping and the sequence of DRAM row activations to minimize the number of DRAM row activations required for a specific operation. 

The third step {(\circled{3} in Figure~\ref{fig_framework_3})} is to program the memory controller to issue the sequence of DRAM row activations to the appropriate rows in DRAM to perform the computation of the operation from start to end. To this end, \mech uses a \emph{control unit} in the memory controller that transparently executes the sequence of DRAM row activations for each specific operation.

\noindent
\textbf{System Integration.} To incorporate SIMDRAM into a real system, we address 
two integration challenges as part of our work:
(1)~managing memory with both vertical and horizontal layouts
in a system, and
(2)~exposing \mech functionality to programmers and compilers.
As part of the support for system integration, we introduce two components.

First, \mech adds a \emph{transposition unit} in the memory controller that transforms the data layout from the conventional horizontal layout to vertical layout (and vice versa), allowing both layouts to coexist. Using the transposition unit, \mech provides the ability to store only the data that is required for in-DRAM computation in the vertical layout. \mech maintains the horizontal layout for the rest of the data and allows the CPU to read/write its operands from/to DRAM in a horizontal layout and at full bandwidth. 
Second, \mech extends the ISA to enable the user/compiler to communicate with the \mech control unit.

\noindent
\textbf{Key Results. }
We demonstrate \mech's  functionality using an example set of operations including  (1)~\emph{N}-input logic operations (e.g., AND/OR/XOR of more than 2 input bits); (2)~relational operations (e.g., equality/inequality check, greater than, maximum, minimum); (3)~arithmetic operations (e.g., addition, subtraction, multiplication, division); (4)~predication (e.g., if-then-else); and (5)~other complex operations such as bitcount and ReLU~\cite{goodfellow2016deep}. 

We compare the benefits of \mech to different state-of-the-art computing platforms (CPU, GPU, and Ambit~\cite{seshadri2017ambit, seshadri2015fast, seshadri2019dram,seshadri2016processing,seshadri.bookchapter17,seshadri.arxiv16}). We comprehensively evaluate SIMDRAM's reliability, area overhead, throughput, and energy efficiency. We leverage the \mech framework to accelerate seven application kernels from machine learning, databases, and image processing (VGG-13~\cite{simonyan2014very}, VGG-16~\cite{simonyan2014very}, LeNET~\cite{lecun2015lenet}, kNN~\cite{lee1991handwritten}, TPC-H~\cite{tpch}, BitWeaving~\cite{li2013bitweaving}, brightness~\cite{gonzales2002digital}). Using a single DRAM bank, SIMDRAM provides
(1)~2.0$\times$ the throughput and 2.6$\times$ the energy efficiency of Ambit~\cite{seshadri2017ambit}, averaged across the 16 implemented operations; and
(2)~2.5$\times$ the performance of Ambit, averaged across the seven application kernels. Compared to a CPU and a high-end GPU, SIMDRAM using 16 DRAM banks provides 
(1)~257$\times$ and 31$\times$ the energy efficiency, and 88$\times$ and 5.8$\times$ the throughput of the CPU and GPU, respectively, averaged across the 16 operations; and 
(2)~21$\times$ and 2.1$\times$ the performance of the CPU and GPU, respectively, averaged across the seven application kernels.  SIMDRAM incurs no additional area overhead on top of Ambit, and a total area overhead of only 0.2\% in a high-end CPU. We also evaluate the reliability of \mech under different degrees of manufacturing process variation, and observe that it guarantees correct operation as the DRAM process technology node scales down to smaller sizes. 

{For more information on our SIMDRAM framework and our extensive evaluation results (including a comparison {to} an alternative framework for processing-using-cache architectures), please refer to our full paper~\cite{hajinazarsimdram,simdramarxiv}.} 

\section{{Discussion}}
\vspace{-5pt}

{Few prior works tackle the challenge of providing end-to-end support for PIM. We describe these works and their limitations for in-DRAM computing.}

{\noindent \textbf{Workload Characterization and Benchmark Suites for PIM.}  
We highlight two prior works, \cite{murphy2001characterization} and PrIM~\cite{gomez2021benchmarking,gomez2021benchmarkingcut,gomez2022benchmarking} that also focus on characterizing workloads and providing benchmark suites for PIM architectures. 
In \cite{murphy2001characterization}, the authors provide the first work that characterizes workloads for PIM. They analyze the benefits a PIM architecture similar to \cite{IRAM_Micro_1997}, where vector processing compute units are integrated into the DDRx memory modules, provides for five applications. Even though \cite{murphy2001characterization} has a similar goal to DAMOV, it understandably does not provide insights into modern data-intensive applications and PIM architectures as it dates from 2001. 
The authors of \cite{gomez2021benchmarking,gomez2021benchmarkingcut,gomez2022benchmarking} propose PrIM, a benchmark suite of 16 workloads from different application domains (e.g., dense/sparse linear algebra, databases, data analytics, graph processing, neural networks, bioinformatics, image processing) tailored to fit the characteristics of a \emph{real} PIM architecture (i.e., the UPMEM-based PIM system~\cite{devaux2019true}). PrIM is open-source and publicly available at~\cite{gomezluna2021repo}.}  {Unlike these prior works, DAMOV is applicable {to} and can be used to study other PIM architectures than processing-in/-near DRAM, including processing-in/-near cache~\cite{eckert2018neural,aga2017compute,dualitycache,denzler2021casper,simon2020blade,nori2021reduct,nag2019gencache}, processing-in/-near storage~{\cite{ghiasi2022genstore,wilkening2021recssd,Cho_2013,riedel2001active,riedel2000active,riedel2000data,riedel1999active,riedel1998active,riedel1997active,keeton1998case,jun2015bluedbm}}, and processing-in/-near emerging NVMs~\cite{pinatubo2016,Shafiee2016,song2017pipelayer,Chi2016,song2018graphr,wang2021rerec,angizi2020exploring}. This is possible since DAMOV's methodology and benchmarks are mainly concerned with broadly characterizing data movement bottlenecks in an application, independent of the underlying PIM architecture.}

{\noindent \textbf{Frameworks for PIM.} 
DualityCache~\cite{dualitycache} is an end-to-end framework for \textbf{in-cache computing}{, which} executes a fixed set of operations in a single-instruction multiple-thread (SIMT) manner. {E}mploying DualityCache in DRAM is not straightforward due to the fundamental differences between in-cache computing and in-DRAM computing (e.g., the destructive behavior of DRAM operations {and cost-sensitivity of DRAM chips}). Two prior works{, Hyper-A~\cite{zha2020hyper} and IMP~\cite{fujiki2018memory}, propose} frameworks for {\textbf{in-emerging-NVM computing}}.
{Since Hyper-A and IMP target in-emerging-NVM substrates that utilize different computing paradigms (e.g., associative processing~\cite{slade1956cryotron,krikelis1994associative}) or rely on particular structures of the NVM array (such as analog-to-digital/digital-to-analog converters) to perform computation, they are \emph{not} applicable to an in-DRAM substrate that performs bulk bitwise operations.} Olgun \textit{et al.} propose the PiDRAM~\cite{olgun2021pidram} framework, a flexible end-to-end and open-source FPGA-based framework that enables system integration studies and evaluation of \textbf{in-DRAM computing} techniques (e.g., in-DRAM copy {and initialization}~\cite{seshadri2013rowclone,seshadri2018rowclone} and in-DRAM true random {generation}~{\cite{kim2019d,bostanci2022dr,olgun2021quac}}) using real unmodified DRAM chips. PiDRAM is publicly available at~\cite{pidram.github} and can be used to prototype our SIMDRAM framework in a real system.}

\section{Conclusion \& Future Work}
\vspace{-5pt}

{This paper summarizes {two of} our recent efforts} {towards} {providing} holistic {system-level} support for processing-in-memory (PIM) systems. We provide 
\emph{(i)} a methodology to identify and characterize sources of data movement bottlenecks in a workload {that can enable the programmer to assess} whether a processing-near-memory (PnM) architecture can mitigate the identified data movement bottlenecks; 
\emph{(ii)} the first benchmark suite {(i.e., \emph{DAMOV})} tailored for {analyzing} data movement bottlenecks {and effects of near-data processing}; and
\emph{(iii)} an end-to-end framework {(i.e., \emph{SIMDRAM})} that enables efficient {and programmer-transparent} computation of a wide range of arbitrary and complex operations {by} employing processing-using-memory (PuM) {in DRAM}. {We believe that DAMOV can enable 
(1) {simple} and practical identification of PIM-suitable workloads {and functions}, 
(2) a research substrate (with our benchmark suite and simulator) for PIM-related architecture {and system} studies{.} SIMDRAM can facilitate the {broader} adoption of PuM architectures {by more workloads and programmers}.} We hope that our work inspires future research on system-level solutions {and tools} that can aid the {research, development, and implementation} of PIM architectures.

\section*{Acknowledgments}
\vspace{-5pt}

We thank the SAFARI Research Group members for valuable feedback and the stimulating intellectual environment they provide. {We acknowledge the generous gifts provided by our industrial partners, including ASML, Facebook, Google, Huawei, Intel, Microsoft, and VMware.
We acknowledge support from the Semiconductor Research Corporation and the 
ETH Future Computing Laboratory.} 

This invited extended abstract is a summary version of our {two} prior works DAMOV~\cite{oliveira2021pimbench,deoliveira2021arxiv} (published at IEEE Access {2021}) and SIMDRAM~\cite{hajinazarsimdram,simdramarxiv} (published at ASPLOS 2021). Presentations that describe DAMOV can {be} found at \cite{damov_talk_short} (short talk video), \cite{damov_talk_long} (long talk video), and \cite{damov_tutorial} (tutorial on the DAMOV framework and benchmarks). {A} presentation that describes SIMDRAM can be found at~\cite{simdram_talk}.

\footnotesize
\bibliographystyle{IEEEtran}
\bibliography{refs}


\end{document}